\documentclass[12pt]{article}
\usepackage{amsmath,epsfig}
\usepackage{amsmath}
\usepackage{amssymb}
\usepackage{mathrsfs}
\usepackage{cite}
\usepackage{amsmath,epsfig}
\usepackage{graphicx}
\usepackage{amsmath,epsfig}
\usepackage{epsfig}
\newsavebox{\PSLASH}
\sbox{\PSLASH}{$p$\hspace{-1.8mm}/}

\begin{document}

\vspace{4cm}
\begin{center}
{\Large\bf{Study of anomalous top quark FCNC interactions via $tW$-channel of single top}}\\
\vspace{1cm} {\bf S. M. Etesami$\
^{\dagger,\ddagger}$,  M. Mohammadi Najafabadi$\
^{\ddagger,}$\footnote{\normalsize{Corresponding author email
address: mojtaba@ipm.ir}}}  \\
\vspace{0.5cm}
{\sl ${\ ^{\dagger}}$ Physics Department,
Isfahan University of Technology (IUT) \\
 Isfahan, Iran}\\
and\\
{\sl ${\ ^{\ddagger}}$  School of Particles and Accelerators, \\
Institute for Research in Fundamental Sciences (IPM) \\
P.O. Box 19395-5531, Tehran, Iran}\\

\vspace{3cm}
 \textbf{Abstract}\\
 \end{center}
The potential of the LHC for investigation of
anomalous top quark interactions with gluon ($tug,tcg$) through
the production of $tW$-channel of single top quark is studied.
In the Standard Model, the single top quarks in the $tW$-channel mode
are charge symmetric meaning that $\sigma(pp\rightarrow t+W^{-}) =
\sigma(pp\rightarrow \bar{t}+W^{+})$. However, the presence of anomalous FCNC
couplings leads to charge asymmetry.
In this paper a method is proposed in which this charge asymmetry may be
used to constrain anomalous FCNC couplings. The strength of resulting
constraints is estimated for the LHC for the center of mass energies
of 7 and 14 TeV.

\newpage

\section{Introduction}

\begin{figure}
\centering
  \includegraphics[width=10cm,height=6cm]{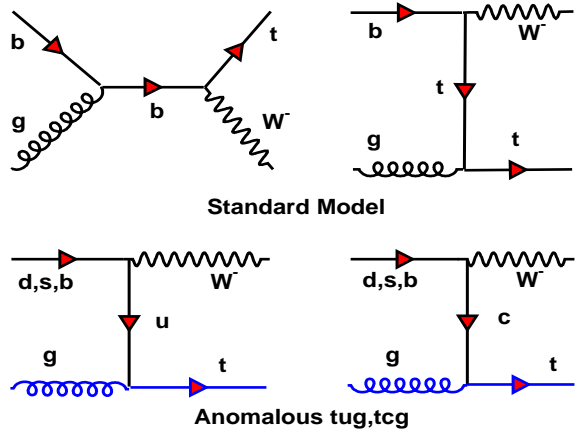}\\
  \caption{Feynman diagrams for the $tW$ channel single top production
at the LHC including anomalous FCNC vertices. }\label{feynman}
\end{figure}

Several properties of the top quark have been measured ever since its
discovery \cite{werner,beneke,tait1,Gerber,
marc,cdftop,d0top}. However, there are still open questions whether
the top quark couplings obey the Standard Model (SM) or there exist
contributions from beyond Standard Model physics.
One tool that is often used to describe the effects of new
physics at an energy scale of $\Lambda$, much higher than the
electroweak scale, is the effective Lagrangian method. If the
underlying extended theory under consideration only becomes
important at a scale of $\Lambda$, then it makes sense to expand the
Lagrangian in powers of $\Lambda^{-1}$ \cite{zhang,han,buchmuller}:
\begin{eqnarray}
\mathcal{L} = \mathcal{L}_{SM} + \sum\frac{c_{i}}{\Lambda^{n_{i}-4}}O_{i}
\end{eqnarray}
where $\mathcal{L}_{SM}$ is the standard model Lagrangian,
$O_{i}$'s are the operators containing {\it only} the SM fields,
$n_{i}$ is the dimension of $O_{i}$ and $c_{i}$'s are
dimensionless parameters.
In the top quark sector, the lowest dimension operators that contribute
to FCNC with the $tcg,tug$ vertex can be written as \cite{beneke}:
\begin{eqnarray}\label{fcnc}
g_{s}\frac{\kappa_{u}}{\Lambda}\bar{u}\sigma^{\mu\nu}
\frac{\lambda^{a}}{2}tG^{a}_{\mu\nu}+h.c.~,~ g_{s}\frac{\kappa_{c}}{\Lambda}\bar{c}\sigma^{\mu\nu}
\frac{\lambda^{a}}{2}tG^{a}_{\mu\nu}+h.c.
\end{eqnarray}
where $g_{s}$ is the strong coupling constant
, $\kappa_{u,c}$ are free
parameters determining the strength of these anomalous couplings and
$G^{a}_{\mu\nu}$ is the gauge field tensor of the gluon. $\lambda_{a}$
are Gell-Mann matrices. $u,c,t$ are Dirac spinors for up,charm and
top quarks and $\sigma_{\mu\nu} = i(\gamma_{\mu}\gamma_{\nu}-\gamma_{\nu}\gamma_{\mu})/2$.
The presence of such anomalous FCNC vertices leads to additional
processes in the $tW$ channel mode of single top production
at hadron colliders such as the LHC.
Figure \ref{feynman} shows the Feynman diagrams
for the production of $tW$ channel of single top in
the SM framework and the new diagrams which are because of the
new anomalous FCNC interactions introduced in Eq.\ref{fcnc}.

Single top quark in the $tW$ mode is not observable at Tevatron
because of its very small cross section. However, at the LHC the
cross section of $tW$ channel at leading order is around 62 pb. It has been
shown that this process is observable at the LHC using the fully simulated data
at the CMS and ATLAS detectors \cite{cmstdr,atlastdr}.
Recently, this process has been studied carefully in \cite{fabio}.

There are many experimental and phenomenological studies about FCNC anomalous
couplings which some can be found in
\cite{tait2,han2,MinYang,aguilar1,aguilar2,aguilar3,
aguilar4,serge1,serge2,tazik,
fcnc1,fcnc2,fcnc3,fcnc4,fcnc5,fcnc6,fcnc7,fcnc8,fcnc9,fcnc10,fcnc11,fcnc12,abazov,cdf,young,cakir
,herq,belyaev,liu,gao1,gao2,drob1,drob2}.
In the SM framework, the $tW$ mode of single top is charge symmetric
meaning that $\sigma(pp\rightarrow t+W^{-}+X) = \sigma(pp\rightarrow \bar{t}+W^{+}+X)$.
The reason is that the parton distribution functions (PDFs) of $b$-quark and
$\bar{b}$-quark in proton are the same.
According to Figure \ref{feynman} in the presence of anomalous couplings
, the $d-$quark contributes to the production
of top quark and $\bar{d}-$quark contributes to the production anti-top quark.
Since the parton distribution function of $d-$quark in the proton is
more than the parton distribution function of $\bar{d}-$quark,
the presence of anomalous FCNC vertices described by Eq.\ref{fcnc} leads to
an asymmetry of charge in the $tW$ channel production. It is worth mentioning
that the charge asymmetry in $tW$-channel can also be generated by non-SM values
of $V_{td}$ and $V_{ts}$ of CKM (Cabibbo-Kobayashi-Maskawa) matrix \cite{agu}.

The aim of this article is to benefit of charge asymmetry to estimate the limits for such anomalous couplings.
Since the two main backgrounds in study of $tW$ channel ($t\bar{t}$, QCD events and $WW$)
are charge symmetric, using charge asymmtery method is considered as a powerful tool
to obtain the limits on anomalous FCNC couplings.

\section{The $tW$-channel cross section and charge asymmetry sensitivities to anomalous couplings}
\begin{figure}
\centering
  \includegraphics[width=7cm,height=7cm]{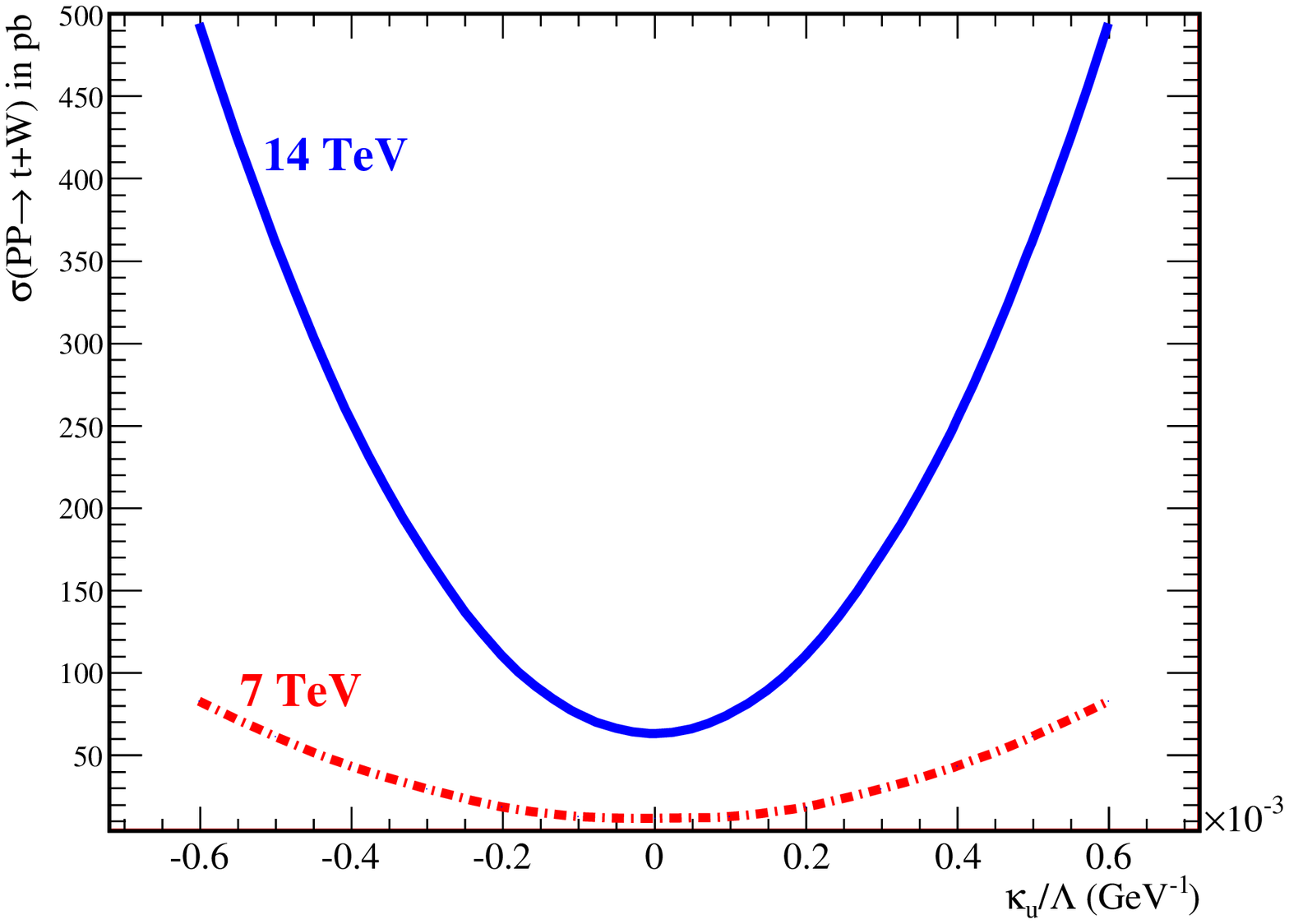}
  \includegraphics[width=7cm,height=7cm]{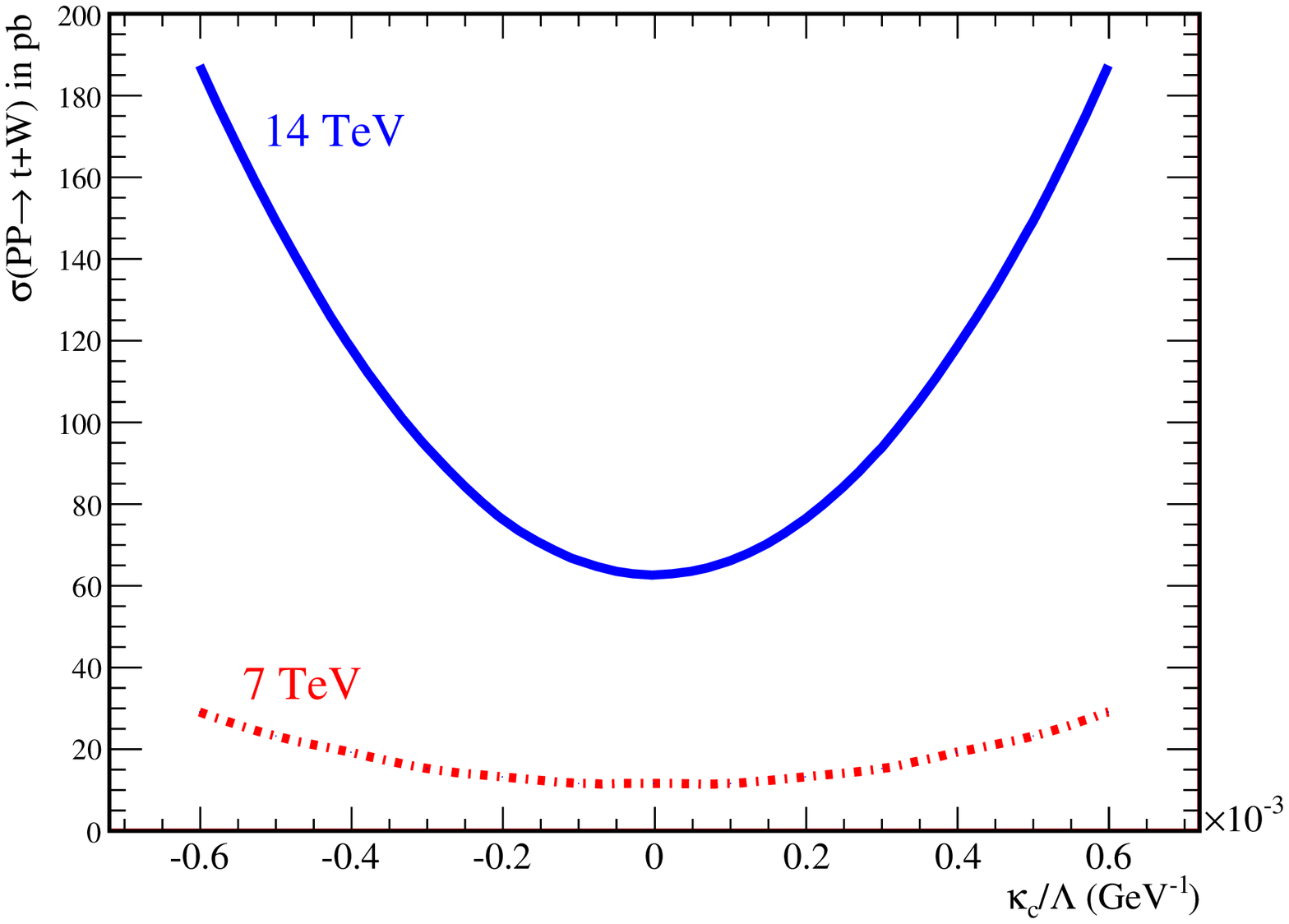}\\
  \caption{The $tW$-cross section dependence on the anomalous couplings at the
LHC with the center of mass energies of 7 TeV and 14 TeV when $\kappa_{c} = 0$ in the left
side and when $\kappa_{u} = 0$ in the right side.}\label{cs}
\end{figure}

The dependency of the $tW$-channel of single top quark cross section
on the anomalous FCNC couplings ($\kappa_{u,c}$) at the LHC with center
of mass energies of 7 TeV and 14 TeV are presented in figure \ref{cs}. This
figure has been obtained using the CompHEP package \cite{comphep}. In
calculation of the cross section, it is assumed that $m_{top} = $
175 GeV/c$^{2}$, $m_{b} = $4.8 GeV/c$^{2}$ and CTEQ6L1 is used as the
proton parton distribution function. The CKM mixing angles are taken as:
$c_{12} = 0.97484$, $c_{23} = 1.0$, $c_{13} = 1.0$.

According to CMS Collaboration full simulation results, the relative
statistical uncertainty on measurement of the cross section
$(\frac{\Delta\sigma}{\sigma})$ of the $tW$-channel taking into account
10 fb$^{-1}$ of integrated luminosity is 9.9$\%$ \cite{cmstdr}. While ATLAS
Collaboration predicted $2.8\%$ for this value with 30 fb$^{-1}$
of integrated luminosity of data \cite{atlastdr}.
Therefore, the cross section of the $tW$ channel will be measured precisely by
the LHC experiments.

In the SM, the cross section of single top quark and single anti-top quark in the
$tW$ channel mode are equal. Therefore:
\begin{eqnarray}
R_{SM} = \frac{\sigma(pp\rightarrow t+W^{-})}{\sigma(pp\rightarrow \bar{t}+W^{+})} = 1.
\end{eqnarray}
However, when the anomalous FCNC vertices are taken into account the above ratio
is not equal to one anymore and $R = R (\kappa_{u},\kappa_{c})$. Figure \ref{chargeratio}
presents the dependency of $R$ on $\kappa_{u}, \kappa_{c}$ at the LHC with the center of mass
 energies of 10 TeV and 14 TeV when $\kappa_{c} = 0$ in the left
side and when $\kappa_{u} = 0$ in the right side.
Due to the higher PDF contributions of the valence quarks w.r.t sea quarks in proton
and the size of the involved CKM matrix elements in the new additional processes in
the production of $tW$ channel single top, $R$ is
more sensitive to $\kappa_{u}$ with respect to $\kappa_{c}$. For example at the
center of mass energy of 14 TeV:
\begin{eqnarray}
R(\kappa_{u}/\Lambda = 0.2~TeV^{-1}, \kappa_{c}/\Lambda  = 0.0) = 1.67\nonumber \\
R(\kappa_{u}/\Lambda  = 0.0, \kappa_{c}/\Lambda  = 0.2~TeV^{-1}) = 1.04
\end{eqnarray}

Therefore, any observable deviation of $R$ from the SM expectation
(charge asymmetry) can be exploited to predict the sensitivity to
anomalous $tug,tcg$ couplings. One should note that the advantage of using
the ratio of $R$ is that the uncertainties coming from
parton distribution function, luminosity and etc. will cancel.

\begin{figure}
\centering
  \includegraphics[width=14cm,height=8cm]{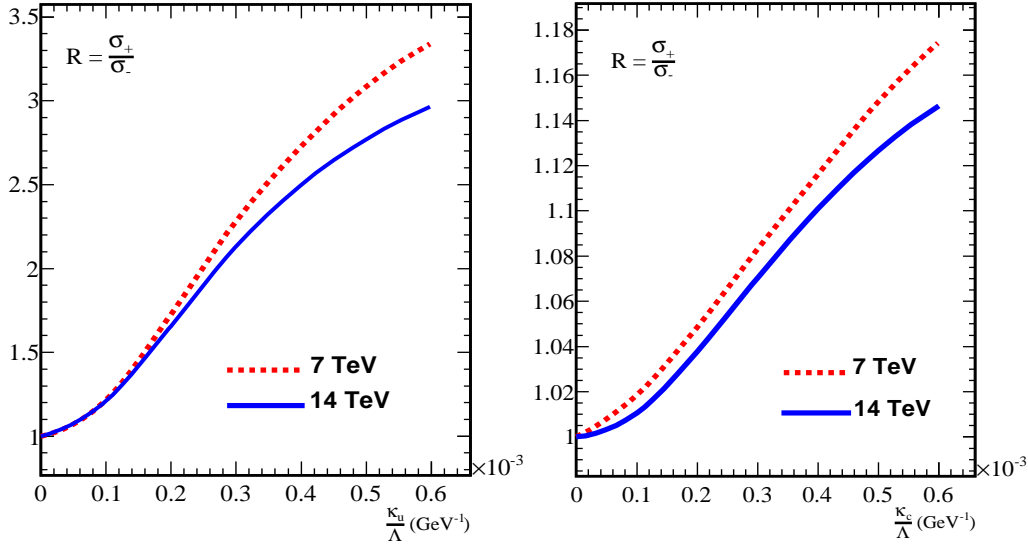}\\
  \caption{The ratio of cross section of top to anti-top in $tW$-channel versus
$\kappa_{u}, \kappa_{c}$ at the LHC with the center of mass
 energies of 7 TeV and 14 TeV when $\kappa_{c} = 0$ in the left
side and when $\kappa_{u} = 0$ in the right side. }\label{ratio}
\end{figure}

\section{Monte Carlo simulation}

In order to predict the sensitivity to the anomalous $tug,tcg$
couplings, we perform Monte Carlo event generation and a very raw detector
simulation (no specific detector is considered). One has to take  into account backgrounds, realistic detector
effects and selection cuts. Obviously, a comprehensive analysis of
all reducible backgrounds and detector effects is beyond
the scope of this study and must be performed by the experimental
collaborations.
In this study the anomalous single top signal events have been
generated by CompHEP package \cite{comphep}.
The CompHEP-PYTHIA interface package \cite{interface} was used to
pass the generated events through PYTHIA \cite{pythia}. PYTHIA performs
fragmentation, parton showering and hadronization.

The detector simulation is performed by smearing energies for
stable particles deposited into proper segmentation of
calorimeter geometry. A jet is clustered by PYCELL routine in
PYTHIA with the cone size of 0.5. B-tagging is simulated with the efficiency
of $60\%$.
The missing transverse energy is calculated by the vector
summation of the lepton and jets.

\section{Event selection and sensitivity study}

In this section after event selection, we predict the
bounds on the anomalous FCNC vertices ($tug,tcg$) using the
{\it semi-leptonic} reconstructed events of $tW$-channel. One
should note that by semi-leptonic we mean that the $W$-boson
coming from the top decays to leptons and another $W$-decays to two jets.
The final state consists of a charged lepton, missing energy,
and three hadronic jets.

To help reduce the backgrounds, we follow the strategy which
ATLAS experiment proposed \cite{beneke},\cite{atlastdr}.
In this strategy one isolated lepton (electron,muon) is required
with transverse momentum\footnote{$p_{T} = \sqrt{p_{x}^{2}+p_{y}^{2}}$}
greater than 20 GeV/c
and $|\eta| < 2.5$\footnote{$\eta = -ln(tan(\frac{\theta}{2}))$}.
The number of jets in the central region ($|\eta| < 2.5$) is
required to be exactly three, each with $p_{T} > 50$ GeV/c.
One of the jets should be tagged as a b-jet. The requirement
of at least one b-jet is necessary to reduce $W+jets$ background events.

To ensure that the other two untagged jets come from the
$W$-boson (which is not from top), it is required that the invariant mass of the
two jets should satisfy: $65~ GeV/c^{2} < m_{jj}< 95~ GeV/c^{2}$.
It is noticeable that this cut and the cut on the number of jets
are very useful to suppress the $W+jets$ background \cite{beneke}.
It is also required that $m_{l\nu b} < 300$ GeV/c$^{2}$ which help suppress
$W+jets$ background. In contrast to
$t\bar{t}$ background, the $W+jets$ background is not charge symmetric.
However, according to the proposed strategy by ATLAS collaboration \cite{beneke},\cite{atlastdr} which
was followed in the current analysis the applied cuts which
mentioned above are powerful in suppressing $W+jets$ background events.
These cuts reduce $W+jets$ background to a negligible level.

Since the charge asymmetry measurement is used in the analysis, the decays of
$tW^{-}\rightarrow W^{+}bW^{-}\rightarrow l^{+}\nu_{l}bjj'$ and
$tW^{-}\rightarrow W^{+}bW^{-}\rightarrow jj'bl^{-}\nu_{l}$ must be
kinematically distinguised. To guarantee that it is required: $m_{bjj'} < 125$ or
$m_{bjj'} > 225$ GeV/c$^{2}$.

The {\it pseudoexperiments} are used for the evaluation of the statistical significance
and including the systematic uncertainties. For the signal process 30,000 random numbers
are drawn from a Gaussian distribution centered on the number of selected events. Further
Gaussian smearing is applied in order to take into account the overall systematic uncertainty.
Calling $G(m,\sigma)$ a random number belonging to a Gaussian distribution with mean $m$ and
standard deviation $\sigma$, each pseudoexperiment gives:
\begin{eqnarray}\label{N}
N^{\pm} = G(N^{\pm}_{sel},\sqrt{N^{\pm}_{sel}})\times G(1,\Delta_{sys}),
\end{eqnarray}
where $N^{\pm}_{sel}$ is the number selected of events after all cuts
with positive and negative charge of the electrons or muons in the top quarks
decay. As discussed before, several uncertainties will cancel when we use
the ratio of $R$ for the analysis. However, few sources of uncertainties may not cancel.
Therefore $\Delta_{sys}$ which is defined as an overall systematic uncertainty
is included in the analysis to get m ore realistic results.

\begin{figure}
\centering
  \includegraphics[width=10cm,height=8cm]{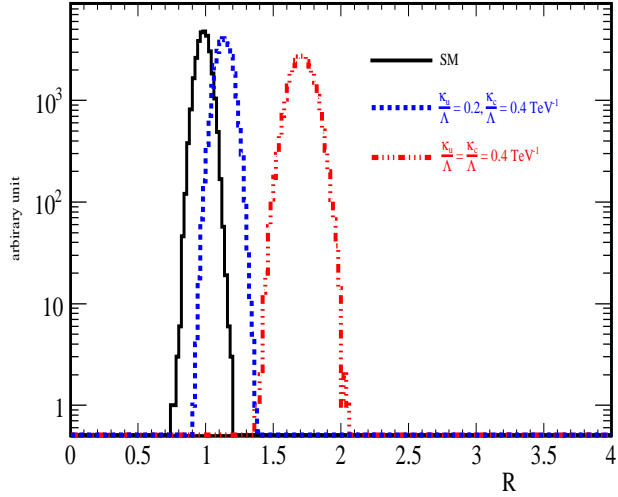}\\
  \caption{The outcome of the pseudoexperiments for $R = \frac{N_{+}}{N_{-}}$ calculated
from Eq.\ref{N}
including $5\%$ systematic uncertainty for the SM and for the presence
of anomalous couplings.}\label{chargeratio}
\end{figure}

Figure \ref{chargeratio} shows the outcome of the pseudoexperiments including $5\%$
systematic uncertainty for $R = \frac{N_{+}}{N_{-}}$ with center of mass energy
of 14 TeV and 10 fb$^{-1}$ of integrated luminosity.  The signal significance is defined as:
\begin{eqnarray}
S = \frac{M(\kappa_{u},\kappa_{c})-M_{SM}}{\sqrt{\sigma^{2}(\kappa_{u},\kappa_{c}) + \sigma^{2}_{SM}}}.
\end{eqnarray}
where $M$ is the peak position and $\sigma$ is the standard deviation of the
distributions. $M$ and $\sigma$ (for the SM case and the presence of anomalous
couplings case) are extracted by Gaussian fits on the pseudoexperiments
distribution in Figure \ref{chargeratio}.
To determine the maximum allowed values of $\frac{\kappa_{u}}{\Lambda}$ and $\frac{\kappa_{c}}{\Lambda}$
that could be reached at the LHC, it is required that $S > 5$ which is
corresponding to approximately $68\%$ confidence level. This
requirement leads to the bounds on $\frac{\kappa_{u}}{\Lambda}$ and $\frac{\kappa_{c}}{\Lambda}$
separately presented in Table \ref{bbxx}. It is noticeable that when the limit on
$\kappa_{u}$ is calculated $\kappa_{c}$ is set to zero and vice versa.

\begin{table}
\begin{center}
\begin{tabular}{|c|c|c|c|}\hline
     & Tevatron& LHC &  LHC  \\
  &        1.96 TeV,2.2fb$^{-1}$ & 7 TeV,1fb$^{-1}$ & 14 TeV,10fb$^{-1}$ \\\hline
 $\kappa_{u}/\Lambda(2\rightarrow 1)$ TeV$^{-1}$ & 0.018 & -    & 0.003\\ \hline
 $\kappa_{u}/\Lambda(2\rightarrow 2)$ TeV$^{-1}$  & 0.037 & -    & 0.006\\ \hline
 $\kappa_{u}/\Lambda(tW)$          TeV$^{-1}$     & -     & 0.1 & 0.08 \\ \hline
 $\kappa_{c}/\Lambda(2\rightarrow 1)$  TeV$^{-1}$ & 0.069 & -    & 0.008\\ \hline
 $\kappa_{c}/\Lambda(2\rightarrow 2)$ TeV$^{-1}$  & 0.15  & -    & 0.013\\ \hline
 $\kappa_{c}/\Lambda(tW)$            TeV$^{-1}$   & -     & 0.38 & 0.35 \\ \hline
\end{tabular} \label{bbxx}
\end{center}\caption{Limits on anomalous couplings obtained from various
experiments and methods.}\label{bbxx}
\end{table}

The FCNC $tqg$-vertex has been studied via other processes such as
quark-gluon fusion process $u(c)+g\rightarrow t$ ($2\rightarrow 1$) or
$qq\rightarrow tq,gg\rightarrow t\bar{q},qg\rightarrow tg$ ($2\rightarrow 2$)
processes. The resulting limits from the studies of $2\rightarrow 1$ and
$2\rightarrow 2$ processes with Tevatron data and LHC simulated data
have been presented in Table \ref{bbxx} \cite{abazov},\cite{cdf}.
One should note that the Tevatron bounds are at $95\%$ confidence level.
The estimated bounds from $2\rightarrow 1$ and $2\rightarrow 2$
are tighter than those obtained in this study. This is because of
the larger cross sections and more statistics of these processes
with respect to the $tW$-channel in the present study.

\section{Conclusion}

The $tW$-channel single top quark production at the LHC was
considered as a probe for non-SM couplings at the LHC. In the SM,
the cross section of single top quark and single anti-top quark
in the $tW$ channel mode are equal. Therefore, $R_{SM} =
\frac{\sigma(pp\rightarrow t+W^{-})}{\sigma(pp\rightarrow
\bar{t}+W^{+})} = 1$. However, when the anomalous FCNC vertices
are taken into account the above ratio is not equal to one
anymore and $R = R (\kappa_{u},\kappa_{c})$. This interesting
aspect was used to extract the $68\%$ C.L. bounds on the
anomalous couplings $\frac{\kappa_{u(c)}}{\Lambda}$. We find that
at 14 TeV center of mass energy and with 10 fb$^{-1}$ integrated
luminosity of data: $\frac{\kappa_{u(c)}}{\Lambda} = 0.08$
TeV$^{-1}$ (0.35). The upper limits for 7 TeV center of mass
energy with 1 fb$^{-1}$ are: $\frac{\kappa_{u(c)}}{\Lambda} = 0.1$
TeV$^{-1}$ (0.38).

\section{Acknowledgment}
The authors would like to thank S. Paktinat and CompHEP authors in particular E.Boos, L. Dudko and A. Sherstnev.


\begin{thebibliography}{99}
\bibitem{werner}   W.~Bernreuther,
  %``Top quark physics at the LHC,''
  J.\ Phys.\ G {\bf 35}, 083001 (2008)
  [arXiv:0805.1333 [hep-ph]].

\bibitem{beneke}  M.~Beneke {\it et al.}, arXiv:hep-ph/0003033.
\bibitem{tait1}  T.~M.~P.~Tait and C.~P.~P.~Yuan,
  %``Single top quark production as a window to physics beyond the Standard
  %Model,''
  Phys.\ Rev.\  D {\bf 63}, 014018 (2001)
  [arXiv:hep-ph/0007298].
\bibitem{Gerber}  C.~E.~Gerber {\it et al.}  [TeV4LHC-Top and Electroweak Working Group],
  %``Tevatron-for-LHC Report: Top and Electroweak Physics,''
  arXiv:0705.3251 [hep-ph].
\bibitem{marc}  M.~A.~Pleier,
  %``Review of Top Quark Properties Measurements at the Tevatron,''
  arXiv:0810.5226 [hep-ex].
\bibitem{cdftop}  T.~Aaltonen {\it et al.}  [CDF Collaboration],
  %``First Observation of Electroweak Single Top Quark Production,''
  Phys.\ Rev.\ Lett.\  {\bf 103}, 092002 (2009)
  [arXiv:0903.0885 [hep-ex]].
\bibitem{d0top}
  V.~M.~Abazov {\it et al.}  [D0 Collaboration],
  %``Observation of Single Top-Quark Production,''
  Phys.\ Rev.\ Lett.\  {\bf 103}, 092001 (2009)
  [arXiv:0903.0850 [hep-ex]].
\bibitem{zhang} R. D. Peccei and X. Zhang, Nucl. Phys. B {\bf 337}, 269 (1990).
\bibitem{han} Z. Han, arXiv:0807.0490[hep-ph].
\bibitem{buchmuller} W.~Buchmuller and D.~Wyler, Nucl.\ Phys.\  B {\bf 268}, 621 (1986).
\bibitem{cmstdr}  G.~L.~Bayatian {\it et al.}  [CMS Collaboration],
  %``CMS technical design report, volume II: Physics performance,''
  J.\ Phys.\ G {\bf 34} (2007) 995.
\bibitem{atlastdr}ATLAS Collaboration, ATLAS Physics TDR Vol. 2, CERN/LHCC/99-15.
\bibitem{fabio} C.~D.~White, S.~Frixione, E.~Laenen and F.~Maltoni,
  %``Isolating Wt production at the LHC,''
  JHEP {\bf 0911}, 074 (2009)
  [arXiv:0908.0631 [hep-ph]].
\bibitem{tait2}  T.~M.~P.~Tait and C.~P.~Yuan,
  %``Anomalous t-c-g coupling: The connection between single top  production and
  %top decay,''
  Phys.\ Rev.\  D {\bf 55}, 7300 (1997)
  [arXiv:hep-ph/9611244].
\bibitem{han2}   T.~Han, M.~Hosch, K.~Whisnant, B.~L.~Young and X.~Zhang,
  %``Single top quark production via FCNC couplings at hadron colliders,''
  Phys.\ Rev.\  D {\bf 58}, 073008 (1998)
  [arXiv:hep-ph/9806486].
\bibitem{MinYang} J. M. Yang, arXiv:0801.0210 [hep-ph].
\bibitem{aguilar1} J.~A.~Aguilar-Saavedra, Acta Phys.\ Polon.\  B {\bf 35}, 2695 (2004).
\bibitem{aguilar2} J.~A.~Aguilar-Saavedra,
  %``A minimal set of top anomalous couplings,''
  Nucl.\ Phys.\  B {\bf 812}, 181 (2009)
  [arXiv:0811.3842 [hep-ph]].
\bibitem{aguilar3} R.~A.~Coimbra, P.~M.~Ferreira, R.~B.~Guedes, O.~Oliveira, A.~Onofre, R.~Santos and M.~Won,
  %``Dimension six FCNC operators and top production at the LHC,''
  Phys.\ Rev.\  D {\bf 79}, 014006 (2009)
  [arXiv:0811.1743 [hep-ph]].
\bibitem{aguilar4}  J.~Carvalho {\it et al.}  [ATLAS Collaboration],
  %``Study of ATLAS sensitivity to FCNC top decays,''
  Eur.\ Phys.\ J.\  C {\bf 52}, 999 (2007)
  [arXiv:0712.1127 [hep-ex]].
\bibitem{serge1} A.~A.~Ashimova and S.~R.~Slabospitsky, Phys.\ Lett.\  B {\bf 668} (2008) 282, arXiv:hep-ph/0604119.
\bibitem{serge2} Yu.~P.~Gouz and S.~R.~Slabospitsky, Phys.\ Lett.\  B {\bf 457} (1999) 177, arXiv:hep-ph/9811330.
\bibitem{tazik}  M.~M.~Najafabadi and N.~Tazik,
  Commun.\ Theor.\ Phys.\  {\bf 52}, 662 (2009)
  [arXiv:0902.0441 [hep-ph]].
\bibitem{fcnc1}
  F.~Larios, R.~Martinez and M.~A.~Perez,
  %``Constraints on top-quark FCNC from electroweak precision measurements,''
  Phys.\ Rev.\  D {\bf 72}, 057504 (2005)
  [arXiv:hep-ph/0412222].
\bibitem{fcnc2}  J.~Abdallah {\it et al.}  [DELPHI Collaboration],
  %``Search for single top production via FCNC at LEP at s**(1/2) = 189-GeV  -
  %208-GeV,''
  Phys.\ Lett.\  B {\bf 590}, 21 (2004)
  [arXiv:hep-ex/0404014].
\bibitem{fcnc3}  G.~A.~Gonzalez-Sprinberg and R.~Martinez,
  %``FCNC top quark decays in extra dimensions,''
  arXiv:hep-ph/0605335.
\bibitem{fcnc4} J.~Cao, Z.~Heng, L.~Wu and J.~M.~Yang,
  %``R-parity violating effects in top quark FCNC productions at LHC,''
  arXiv:0812.1698 [hep-ph].
\bibitem{fcnc5} J.~A.~Aguilar-Saavedra and B.~M.~Nobre,
  %``Rare top decays t --> c gamma, t --> c g and CKM unitarity,''
  Phys.\ Lett.\  B {\bf 553}, 251 (2003)
  [arXiv:hep-ph/0210360].
\bibitem{fcnc6}  S.~Bejar, J.~Guasch, D.~Lopez-Val and J.~Sola,
  %``FCNC-induced heavy-quark events at the LHC from Supersymmetry,''
  Phys.\ Lett.\  B {\bf 668}, 364 (2008)
  [arXiv:0805.0973 [hep-ph]].
\bibitem{fcnc7} S.~Bejar, J.~Guasch and J.~Sola,
  %``Production and FCNC decay of supersymmetric Higgs bosons into heavy  quarks
  %in the LHC,''
  JHEP {\bf 0510}, 113 (2005)
  [arXiv:hep-ph/0508043].
\bibitem{fcnc8} J.~A.~Aguilar-Saavedar,[arXiv:1003.3173[hep-ph]].
\bibitem{fcnc9} G. Eilam, J. L. Hewett and A. Soni, Phys. Rev. D {\bf 44} (1991) 1473 [Erratum-ibid.
D 59 (1999) 039901].
\bibitem{fcnc10} B. Mele, S. Petrarca and A. Soddu, Phys. Lett. B {\bf 435} (1998) 401
[hep-ph/9805498].
\bibitem{fcnc11} J. A. Aguilar-Saavedra and B. M. Nobre, Phys. Lett. B {\bf 553} (2003) 251
[arXiv:hep-ph/0210360]; F. del Aguila, J.A. Aguilar-Saavedra and L. Ametller, Phys. Lett. B {\bf 462} (1999) 310 [arXiv:hep-ph/9906462];
F. del Aguilar and J. A. Aguilar Saavedra, Nucl. Phys. B {\bf 576} (2000) 56 [arXiv:hep-ph/9909222].
\bibitem{fcnc12}P. M. Ferreira, O. Oliveira and R. Santos, Phys. Rev. D {\bf 73} (2006) 034011
[arXiv:hep-ph/0510087]; P. M. Ferreira and R. Santos, Phys. Rev. D {\bf 74 }(2006) 014006 [arXiv:hep-ph/0604144];
P. M. Ferreira and R. Santos, Phys. Rev. D {\bf 73} (2006) 054025 [arXiv:hep-ph/0601078].
\bibitem{abazov}
  V.~M.~Abazov {\it et al.}  [D0 Collaboration],
  %``Search for production of single top quarks via flavor-changing neutral
  %currents at the Tevatron,''
  Phys.\ Rev.\ Lett.\  {\bf 99}, 191802 (2007)
  [arXiv:hep-ex/0702005].
\bibitem{cdf}  T.~Aaltonen {\it et al.}  [CDF Collaboration],
  %``Search for top-quark production via flavor-changing neutral currents in W+1
  %jet events at CDF,''
  arXiv:0812.3400 [hep-ex].
\bibitem{young}  M.~Hosch, K.~Whisnant and B.~L.~Young,
  %``Direct top quark production at hadron colliders as a probe of new
  %physics,''
  Phys.\ Rev.\  D {\bf 56}, 5725 (1997)
  [arXiv:hep-ph/9703450].
\bibitem{cakir}  O.~Cakir and S.~A.~Cetin,
  %``Anomalous Single Top Quark Production At The Cern Lhc,''
  J.\ Phys.\ G {\bf 31}, N1 (2005).
\bibitem{herq}
  M.~Herquet, R.~Knegjens and E.~Laenen,
  %``Single top production in a non-minimal supersymmetric model,''
  arXiv:1005.2900 [hep-ph].
\bibitem{belyaev}
  N.~Kidonakis and A.~Belyaev,
  %``FCNC top quark production via anomalous tqV couplings beyond leading
  %order,''
  JHEP {\bf 0312}, 004 (2003)
  [arXiv:hep-ph/0310299].
\bibitem{liu}
  J.~J.~Liu, C.~S.~Li, L.~L.~Yang and L.~G.~Jin,
  %``Next-to-leading order QCD corrections to the direct top quark  production
  %via model-independent FCNC couplings at hadron colliders,''
  Phys.\ Rev.\  D {\bf 72}, 074018 (2005)
  [arXiv:hep-ph/0508016].
\bibitem{gao1}
 J.~J.~Zhang, C.~S.~Li, J.~Gao, H.~Zhang, Z.~Li, C.~P.~Yuan and T.~C.~Yuan,
  %``Next-to-leading order QCD corrections to the top quark decay via
  %model-independent FCNC couplings,''
  Phys.\ Rev.\ Lett.\  {\bf 102}, 072001 (2009)
  [arXiv:0810.3889 [hep-ph]].
\bibitem{gao2}
  J.~J.~Zhang, C.~S.~Li, J.~Gao, H.~X.~Zhu, C.~P.~Yuan and T.~C.~Yuan,
  %``Next-to-leading order QCD corrections to the top quark decay via the
  %Flavor-Changing Neutral-Current operators with mixing effects,''
  arXiv:1004.0898 [hep-ph].
\bibitem{drob1}
  J.~Drobnak, S.~Fajfer and J.~F.~Kamenik,
  %``Flavor Changing Neutral Coupling Mediated Radiative Top Quark Decays at
  %Next-to-Leading Order in QCD,''
  arXiv:1004.0620 [hep-ph].
\bibitem{drob2}
  J.~Drobnak, S.~Fajfer and J.~F.~Kamenik,
  %``Signatures of NP models in top FCNC decay t --> c(u) l+ l-,''
  JHEP {\bf 0903}, 077 (2009)
  [arXiv:0812.0294 [hep-ph]].
\bibitem{agu} J.~A.~Aguilar-Saavedar, A. Onofre, [arXiv:1002.4718[hep-ph]].
\bibitem{comphep}  E.~Boos {\it et al.}  [CompHEP Collaboration],
  %``CompHEP 4.4: Automatic computations from Lagrangians to events,''
  Nucl.\ Instrum.\ Meth.\  A {\bf 534}, 250 (2004)
  [arXiv:hep-ph/0403113].
\bibitem{interface}  A.~S.~Belyaev {\it et al.},
  %``CompHEP-PYTHIA interface: Integrated package for the collision events
  %generation based on exact matrix elements,''
  arXiv:hep-ph/0101232.
\bibitem{pythia}  T.~Sjostrand, S.~Mrenna and P.~Skands,
  %``PYTHIA 6.4 physics and manual,''
  JHEP {\bf 0605}, 026 (2006)
  [arXiv:hep-ph/0603175].
\bibitem{ptdr2}  G.~L.~Bayatian {\it et al.}  [CMS Collaboration],
  %``CMS technical design report, volume II: Physics performance,''
  J.\ Phys.\ G {\bf 34}, 995 (2007).
\bibitem{wpol}   F.~Hubaut, E.~Monnier, P.~Pralavorio, K.~Smolek and V.~Simak,
  %``ATLAS sensitivity to top quark and W boson polarization in t anti-t
  %events,''
  Eur.\ Phys.\ J.\  C {\bf 44S2}, 13 (2005)
  [arXiv:hep-ex/0508061].
\bibitem{onofre} J.~A.~Aguilar-Saavedra, J.~Carvalho, N.~F.~Castro, A.~Onofre and F.~Veloso,
  %``ATLAS sensitivity to Wtb anomalous couplings in top quark decays,''
  Eur.\ Phys.\ J.\  C {\bf 53}, 689 (2008)
  [arXiv:0705.3041 [hep-ph]].
\bibitem{pdg} C. Amsler et al., Phys.\ Lett.\ B {\bf 667} (2008) 1.
\end{thebibliography}
\end{document}